\documentclass[11pt]{article}
\usepackage[numbers]{natbib}
\usepackage{rotating}
\usepackage{graphicx}

\begin{document}
\def\smallfrac#1#2{\hbox{${{#1}\over {#2}}$}}
\newcommand{\MS}{$\overline{\textrm{MS}}$}

\begin{flushright}
Edinburgh 2016/21\\
\end{flushright}
\vspace{1cm}

\begin{center}
{\Large \bf Charm Production: Pole Mass or Running Mass?}
\vspace{1cm}

Richard~D.~Ball

\vspace{.3cm}
{\it The Higgs Centre for Theoretical Physics, University of Edinburgh,\\
JCMB, KB, Mayfield Rd, Edinburgh EH9 3JZ, Scotland\\
}
\end{center}

\vspace{1cm}

\begin{center}
  {\bf \large Abstract}
\end{center}
  
We consider the role of the charm mass in PDF fits, and in particular the reliability of recent determinations of the \MS\ running mass from fits to inclusive charm production data. We explain how the use of the pole mass, with a charm PDF determined empirically, can reduce PDF uncertainties from charm mass effects, and thereby lead to more reliable predictions for high energy processes at the LHC.

\clearpage

\section*{Introduction}

Inclusive processes involving massive quarks are an important ingredient of LHC physics, not least because of their role in determining PDFs \cite{Rojo:2015acz,Ball:2015oha}. Uncertainties in heavy quark masses can lead to substantial contributions to PDF uncertainties, and thus to uncertainties in predictions for LHC cross-sections for a wide range of processes. 

Perturbative coefficient functions can be renormalized to depend on either the pole mass 
$m$ or the \MS\ running mass $\overline{m}(\mu)$. The perturbative relation between them is now known to four loops \cite{Chetyrkin:1999qi,Marquard:2015qpa}, and for top the choice is essentially immaterial. However for charm and beauty nonperturbative corrections are more substantial, and while the \MS\ charm and beauty masses can be determined rather precisely (to a few tens of MeV) through nonperturbative lattice or sum rule calculations \cite{Erler:2014eya}, their pole masses are subject to large uncertainties (a few hundreds of MeV).

Global PDF fits \cite{Ball:2014uwa,Harland-Lang:2014zoa,Dulat:2015mca} traditionally use pole masses. This is because the experimental observables used in the fits are generally inclusive, to avoid large uncertainties from final state effects. Heavy quarks in the final state are dealt with in the perturbative calculations by putting them on-shell: hadronisation corrections are then of relative order $\Lambda/m$, and thus power suppressed. The on-shell condition naturally leads to the mass dependence from heavy quarks in the final state being expressed in terms of the pole mass. 

In this short note, we will re-examine the possible use of the \MS\ running mass in PDF determinations \cite{Alekhin:2012vu,Gao:2013dva,Cooper-Sarkar:2013aca,Abramowicz:2014zub},
and consider other ways in which uncertainties due to charm mass dependence might be reduced at LHC \cite{Ball:2016neh}.

\section*{Pole Masses \& Running Masses}

The pole mass $m$ and the \MS\ mass at scale $\mu$, $\overline{m}(\mu)$ are related through perturbative expansion: 
\begin{equation}
m^2 =\overline{m}(\mu)^2(1 + z(\alpha_s)) = \overline{m}(\mu)^2(1 + \sum_1^\infty z_n(\mu)\alpha_s(\mu)^n),
\label{eq:polemsbar}
\end{equation}
where the perturbative coefficients $z_n(\mu)$ are known to four loops \cite{Chetyrkin:1999qi,Marquard:2015qpa}. Choosing $\mu=\overline{m}(\mu)$ consistently in both mass and coupling,
\begin{eqnarray}
m_t &=& \overline{m}_t(1+0.424\alpha_s + 0.835\alpha_s^2 + 2.38\alpha_s^3 + 8.62\alpha_s^4 + O(\alpha_s^5)),\label{eq:mtexp}\\
m_b &=& \overline{m}_b(1+0.424\alpha_s + 0.940\alpha_s^2 + 3.05\alpha_s^3 + 12.7\alpha_s^4 + O(\alpha_s^5)),\label{eq:mbexp}\\
m_c &=& \overline{m}_c(1+0.424\alpha_s + 1.046\alpha_s^2 + 3.76\alpha_s^3 + 17.5\alpha_s^4 + O(\alpha_s^5)),\label{eq:mcexp}
\end{eqnarray}
whence (assuming $\alpha_s(m_Z)=0.118$, and evolving to $\mu=\overline{m}(\mu))$
\begin{eqnarray}
m_t 
&=& \overline{m}_t(1+0.046 + 0.010 + 0.003 + 0.001 + \cdots),\label{eq:mtexpn}\\
m_b 
&=& \overline{m}_b(1+0.096 + 0.048 + 0.035 + 0.033 + \cdots),\label{eq:mbexpn}\\
m_c 
&=& \overline{m}_c(1+0.16 + 0.16 + 0.22 + 0.39 + \cdots).\label{eq:mcexpn}
\end{eqnarray}
It is clear from these numbers that while the top mass expansion is converging well (at least up to four loops), the bottom mass expansion may be starting to break down after three loops, and the charm mass expansion fails already at two loops. This breakdown is due to renormalons: indeed it can be shown \cite{Beneke:2016cbu} that the ultimate uncertainty in the top quark pole mass from this source is less than one part per mille. For beauty, and in particular for charm, the situation is much worse: as can be seen from inspection of the series, the uncertainty for beauty is of the order of a few per cent, while for charm it is a few tens of per cent. This is roughly consistent with a naive estimate of nonperturbative corrections of the order of $\Lambda/m$ in each case, with $\Lambda$ a typical QCD scale of order a few hundred MeV.

\section*{Kinematics of Charm Production}

Consider for definiteness a deep inelastic process, with production of charm quarks in the final state. 
Adopting a frame in which the 4-vectors $q$ and $p$ are collinear, we can write $p^{\mu} = \big(p^+,\smallfrac{M^2}{p^+},\mbox{\bf{0}}\big)$, $q^{\mu} = -\big(\eta p^+,\smallfrac{-Q^2}{\eta p^+},\bf{0}\big)$, where $M$ is the mass of the target nucleon (so $p^2=M^2$, $q^2=-Q^2$). Then $2q.p = Q^2/x = Q^2/\eta - \eta M^2$, so
\begin{equation}
\eta = 2x/(1+\sqrt{1+\smallfrac{4x^2M^2}{Q^2}})
\label{eq:tmc}
\end{equation}
is Bjorken-x corrected for target mass effects. 
Note that $\eta < x$ so including the target mass increases the phase space for the process.

Now consider a hard process in which heavy quarks are produced: the paradigm processes are photon-gluon fusion, which begins at $O(\alpha_s)$ through the diagram 
$\gamma^* g\to c\overline{c}$, and dimuon production, which begins at $O(1)$ through the conversion 
$W^* s\to c$. Assuming that the incoming parton is massless, then we have (in a collinear frame) $k^{\mu} = \big(\xi p^+,0,\mbox{\bf{0}}\big)$:
the hard process then requires $\hat{s} = (q+k)^2 = (\smallfrac{\xi}{\eta}-1)Q^2 \geq \hat{s}_{\rm th}$, where $\hat{s}_{\rm th}$ is the threshold energy, and thus that $\xi\geq\eta (1+\smallfrac{\hat{s}_{\rm th}}{Q^2})$. The target remnant $X$ will then have 4-momentum $p_X = p-k$, and $p_X^2 = (1-\xi)M^2 >0$ means that $\xi \leq 1$. Thus the momentum fraction of the incoming parton must lie in the interval
\begin{equation}
\chi\equiv \eta (1+\smallfrac{\hat{s}_{\rm th}}{Q^2}) \leq \xi \leq 1.
\label{eq:chi}
\end{equation}
The dominant effect of the heavy quarks in the final state is thus to reduce the phase space available in the hard process. 

Looking at the process in its entirity (ie including the final state $X$), to be above threshold we must have $W^2 = (q+ p)^2 \geq W_{\rm th}$, where $W_{\rm th}\geq \hat{s}_{\rm th}$ is the threshold energy for the entire process: this means that 
\begin{equation}
x \leq 1/(1+\smallfrac{W_{\rm th}}{Q^2})\equiv x_{\rm max}.
\label{eq:xmaxnc}
\end{equation} 
This is consistent with Eq.(\ref{eq:chi}) since it ensures that $\chi< x/x_{\rm max} < 1$ provided $W_{\rm th}\geq \hat{s}_{\rm th}\geq \hat{s}_{\rm th}$.   

The \MS\ factorization of the structure function $F\big(x;\smallfrac{m_c^2}{Q^2}\big)$ into a convolution of a hard coefficient function with the light parton PDF $f(\xi,\mu^2)$ may now be written in the form 
\begin{equation}
F\big(x;\smallfrac{m_c^2}{Q^2}\big) = \int_{\chi}^1 \frac{d\xi}{\xi} C\big(\xi;\smallfrac{Q^2}{\mu^2},\smallfrac{m_c^2}{Q^2}\big)f\big(\smallfrac{\chi}{\xi},\mu^2\big) 
+ O\big(\smallfrac{\Lambda}{Q}\big), 
\label{eq:conv}
\end{equation} 
where $\mu$ is the factorization scale. Note that this is equivalent to the commonly used form 
\begin{equation}
\int_{\eta}^1 \frac{dy}{y} \theta(\hat{s}-\hat{s}_{\rm th})\tilde{C}\big(y;\smallfrac{Q^2}{\mu^2},\smallfrac{m_c^2}{Q^2}\big)f\big(\smallfrac{\eta}{y},\mu^2\big)
= \int_{\eta}^{x_{\rm max}} \frac{dy}{y} \tilde{C}\big(y;\smallfrac{Q^2}{\mu^2},\smallfrac{m_c^2}{Q^2}\big)f\big(\smallfrac{\eta}{y},\mu^2\big),
\label{eq:convth}
\end{equation}
as may be readily seen by rescaling the integration variable: $\xi = y/x_{\rm max}$, $\eta/x_{\rm max} = \chi$, and thus for equivalence 
$C(\xi;\smallfrac{Q^2}{\mu^2},\smallfrac{m_c^2}{Q^2}) = \tilde{C}(x_{\rm max}\xi;\smallfrac{Q^2}{\mu^2},\smallfrac{m_c^2}{Q^2})$. The coefficient function 
$C(\xi;\smallfrac{Q^2}{\mu^2},\smallfrac{m_c^2}{Q^2})$ may be computed in a FFNS, or some form of VFNS or S-VFNS by adding to it a resummed massless calculation and subtracting the double counting 
\cite{Cacciari:1998it,Collins:1998rz,Kramer:2000hn,Forte:2010ta,Guzzi:2011ew,Ball:2015dpa}
--- this is preferable since it offers a more accurate description of processes with $Q^2\gg m_c^2$ \cite{Ball:2013gsa}. However the precise details are not necessary for the following, since the mass dependence near threshold remains the same in each case.

The convolution in Equation~(\ref{eq:conv}) may be undone by defining Mellin moments in the usual way: if
\begin{equation}
C(N;\smallfrac{Q^2}{\mu^2},\smallfrac{m_c^2}{Q^2}) \equiv \int_{0}^1 \frac{d\xi}{\xi} \xi^N C(\xi;\smallfrac{Q^2}{\mu^2},\smallfrac{m_c^2}{Q^2}),\qquad 
f(N;\mu^2) \equiv \int_{0}^1 \frac{d\xi}{\xi} \xi^N f(\xi,\mu^2),
\label{eq:melldef}
\end{equation}
then
\begin{eqnarray}
F\big(x,\smallfrac{m_c^2}{Q^2}\big) &=& \int_C\frac{dN}{2\pi i}\, \chi^{-N} C(N;\smallfrac{Q^2}{\mu^2},\smallfrac{m_c^2}{Q^2})f(N;\mu^2)+ O\big(\smallfrac{\Lambda}{Q}\big)\nonumber\\
&=& \int_C\frac{dN}{2\pi i} \,\eta^{-N} \Big(1+\smallfrac{\hat{s}_{\rm th}}{Q^2}\Big)^{-N} C(N;\smallfrac{Q^2}{\mu^2},\smallfrac{m_c^2}{Q^2})f(N;\mu^2)+ O\big(\smallfrac{\Lambda}{Q}\big), 
\label{eq:mellcon}
\end{eqnarray}
with $C$ the usual contour for Mellin inversion. For $\chi < 1$ we close the contour on the left, and pick up the residues of the poles in the integrand: for $\chi > 1$ we close on the right and get zero. Note that the factor of 
$(1+\smallfrac{\hat{s}_{\rm th}}{Q^2})^{-N}$ is needed to get the thresholds right: the essential singularity at $N=\infty$ in this factor produces the theta-function in Equation~(\ref{eq:convth}). 

The choice of threshold $\hat{s}_{\rm th}$ will depend on the production process under consideration: for example in semi-inclusive $J/\psi$ production $\hat{s}_{\rm th}=m_J^2$, while for charged current semi-inclusive $D$-meson production $\hat{s}_{\rm th}=m_D^2$. For such processes the only explicit charm mass dependence in Equation~(\ref{eq:mellcon}) is through internal heavy quark propagators in the hard coefficient function. Changing the definition of the charm mass in the coefficient function from pole to running is then straightforward \cite{Alekhin:2010sv}, since they are analytic in $m_c^2/Q^2$: we can determine 
\begin{equation}
\overline{C}(N;\smallfrac{Q^2}{\mu^2},\smallfrac{\overline{m}_c^2}{Q^2})=C(N;\smallfrac{Q^2}{\mu^2},\smallfrac{m_c^2}{Q^2}),
\label{eq:msbarcf}
\end{equation}
order by order in perturbation theory by substituting Equation~(\ref{eq:polemsbar}) in the r.h.s. and expanding everything in powers of $z(\alpha_s)$, and thus $\alpha_s$. It is easy to see that the result is precisely what we would get if we had renormalized the charm mass in \MS\ rather than the pole scheme from the start, and that the singularity structure of the coefficient functions in Mellin space, and in particular their behaviour at large $N$ due to soft gluon emission(see for example \cite{Laenen:1998kp,Corcella:2003ib} and ref therein), will be unchanged. 

\section*{Inclusive Charm Production}

When determining PDFs, semi-inclusive processes are not so useful, since they have uncontrolled theoretical uncertainties due to the fragmentation functions. For inclusive charm production, it is usual to assume that the heavy quark goes on-shell in the final state, and this introduces an additional (purely kinematic) dependence on the charm pole mass. For example, for the inclusive neutral current process $\hat{s}_{\rm th}=4m_c^2$, where $m_c$ is the pole mass of the charm quark. Of course there will also be hadronisation corrections, but the uncertainty due to these can be reasonably argued to be modelled through a shift of $m_c$ by $\Lambda$: then the threshold factor in Equation~(\ref{eq:mellcon})
\begin{eqnarray}
\Big(1+\smallfrac{4m_c^2}{Q^2}\Big)^{-N} &\to& \Big(1+\smallfrac{4{m}_c^2}{Q^2}\Big)^{-N}\Big(1 + \smallfrac{4(\Lambda(2 m_c+\Lambda)}{Q^2+4{m}_c^2}\Big)^{-N}\nonumber\\
&=& \Big(1+\smallfrac{4{m}_c^2}{Q^2}\Big)^{-N}\Big[1 +O \Big(\smallfrac{\Lambda m_c}{Q^2}\Big)\Big],
\end{eqnarray}
so close to threshold the correction is the same size as the higher twist terms already neglected. It follows that, while inclusive processes are most naturally computed using the pole mass, at leading twist we only need to know the value of the pole mass up to terms of $O(\Lambda)$. Naturally this coincides with the limitation on the precision of $m_c$ due to the poor convergence of Equation~(\ref{eq:mcexp}).

The inclusive structure function Equation~(\ref{eq:mellcon}) now depends explicitly on the charm pole mass in two ways: through internal charm quark lines, expressed through the dependence of the hard coefficient function $C(N;\smallfrac{Q^2}{\mu^2},\smallfrac{m_c^2}{Q^2})$, but additionally through the final state charm quark lines going on-shell, and giving rise to the explicit factor $(1+4m_c^2/Q^2)^{-N}$ which imposes the correct threshold kinematics. If we want to switch from a pole mass calculation to a calculation expressed entirely in terms of the \MS\ mass, we must convert not only the coefficient functions, using Equation~(\ref{eq:msbarcf}), but also the threshold factor $(1+4m_c^2/Q^2)^{-N}$: using Equation~(\ref{eq:polemsbar})
\begin{equation}
\Big(1+\smallfrac{4m_c^2}{Q^2}\Big)^{-N} = \Big(1+\smallfrac{4\overline{m}_c^2}{Q^2}\Big)^{-N}\Big(1 + \smallfrac{4\overline{m}_c^2}{Q^2+4\overline{m}_c^2}z(\alpha_s)\Big)^{-N}
\end{equation}
While the first factor on the r.h.s. now moves the threshold from $4m_c^2$ to $4\overline{m}_c^2$, as required, the second is a perturbative expansion in powers of $\alpha_s$ which can be absorbed into corrections to the coefficient function, which then becomes
\begin{eqnarray}
&&\overline{C}(N;\smallfrac{Q^2}{\mu^2},\smallfrac{\overline{m}_c^2}{Q^2})\exp\big(-N\ln(1 + \smallfrac{4\overline{m}_c^2}{Q^2+4\overline{m}_c^2}z(\alpha_s))\big)\nonumber\\\qquad\qquad &=&
\overline{C}(N;\smallfrac{Q^2}{\mu^2},\smallfrac{\overline{m}_c^2}{Q^2})\Big[1 - \smallfrac{4\overline{m}_c^2}{Q^2+4\overline{m}_c^2} N z(\alpha_s)+O(z^2N^2)\Big].
\end{eqnarray}
However because the threshold factor has an essential singularity at $N=\infty$, these corrections behave at large $N$ as $\alpha_s^nN^{n}$, i.e. they grow much more strongly than the \MS\ threshold logarithms $\alpha_s^n(\ln N)^{2n-1}$ in the pole mass coefficient function $C(N;\smallfrac{Q^2}{\mu^2},\smallfrac{m_c^2}{Q^2})$ \cite{Laenen:1998kp,Corcella:2003ib}. This spoils the perturbative expansion at large $N$, i.e. in the threshold region, by altering the form of the singularity at $N=\infty$. 

The same problem is evident in the convolution Equation~(\ref{eq:conv}): a factor of $N$ in Mellin space (\ref{eq:mellcon}) corresponds to a derivative of $\xi$ in Equation~(\ref{eq:conv}), and indeed expanding out the mass dependence in $\chi$ results in terms involving derivatives of the PDF with respect to $\xi$ \cite{Alekhin:2010sv}, which spoil the \MS\ factorization of the perturbative expansion. Integrating by parts, to take these derivatives off the PDF and onto the coefficient function, generates instead the spurious soft singularities in the coefficient function (as extra factors of 
$\alpha_s^n/(1-\xi)^{n+1}$). Note that far above threshold the extra terms are suppressed by $(\overline{m}_c^2/Q^2)^{n}$: however close to threshold they ruin the perturbative expansion.

It follows that we can only compute inclusive processes reliably using \MS\ masses to a particular order in perturbation theory if we convert only the coefficient functions (using Eq.~(\ref{eq:msbarcf})), but leave the threshold factor in terms of the pole mass. The dependence of the threshold on the pole mass then in effect resums the large $N$ power corrections, of order $\alpha_s^nN^{n}$, which would otherwise spoil the perturbative expansion. However this effectively negates any advantage that might be gained from using the precise \MS\ masses determined nonperturbatively, since Equation~(\ref{eq:mcexp}) expressing the pole mass in terms of the \MS\ mass already has an intrinsic uncertainty of a few hundred MeV. For inclusive charm production there thus seems little practical alternative to using the pole mass throughout, and determining it empirically.

\section*{Conclusions}

We have shown that, while for processes with only internal charm quark lines (such as processes with no charm in the final state, or semi-inclusive processes) it is straightforward to calculate using either pole mass or running mass in the hard cross-section, for inclusive processes with charm in the final state there no advantage to using the running mass, since the kinematics produces a nontrivial dependence on the pole mass which cannot be avoided without spoiling the factorized perturbative expansion. It follows that any empirical determination of the charm quark mass from inclusive charm production data has an intrinsic limitation due to nonperturbative corrections of a few hundred MeV. It is easy to see that these considerations generalise straightforwardly to inclusive hadronic processes such as $Wc$ or $Z c\overline{c}$ production, and indeed to inclusive beauty production, though here the effect will be less significant. 

A number of recent perturbative determinations of the \MS\ charm mass from inclusive data \cite{Alekhin:2012vu,Gao:2013dva,Cooper-Sarkar:2013aca,Abramowicz:2014zub} claim an uncertainty as small as $50$ MeV, competitive with the nonperturbative results \cite{Erler:2014eya}. The reason for this small uncertainty is probably the use of the theoretical assumption that charm is produced entirely perturbatively, which greatly increases sensitivity to the charm mass. However it is clear from the poor convergence of Equation~(\ref{eq:mcexp}) that perturbation theory close to the charm threshold is unreliable: charm production is subject to large nonperturbative corrections. Relaxing the assumption by fitting a charm PDF 
\cite{Ball:2016neh,Ball:2015tna}, significantly reduces the dependence of the PDFs on the charm mass. This in turn reduces the dependence of high energy cross-sections required at LHC on the charm mass, and thus increases their precision \cite{Ball:2016neh}. It will also presumably increase the uncertainty on any empirical determination of the charm mass from inclusive data to a few hundred MeV. We intend to check this in a future NNPDF study.

\section*{Acknowledgements}
The author would like to thank Valerio Bertone for discussions on these issues, and the organisers of Diffraction16 for their hospitality in Sicily.



\end{document}